\documentclass[aps,prd,onecolumn,superscriptaddress,nofootinbib,showpacs]{revtex4}

\usepackage{graphicx}
\usepackage{slashed}
\usepackage{amsmath,bbm,latexsym,amssymb}
\usepackage{epsfig}
\usepackage{epstopdf}

\newcommand{\be}{\begin{equation}}
\newcommand{\ee}{\end{equation}}
\newcommand{\ba}{\begin{eqnarray}}
\newcommand{\ea}{\end{eqnarray}}
\newcommand{\bs}{\begin{subequations}}
\newcommand{\es}{\end{subequations}}

\newcommand{\grts}{\raise.3ex\hbox{$>$\kern-.75em\lower1ex\hbox{$\sim$}}}
\newcommand{\lets}{\raise.3ex\hbox{$<$\kern-.75em\lower1ex\hbox{$\sim$}}}

\begin{document}
\vspace*{1cm}

\title{Probing the scalar-pseudoscalar mixing in the 125 GeV Higgs particle with current data}

\author{A.\ Barroso}
\affiliation{Centro de F\'{\i}sica Te\'{o}rica e Computacional,
    Faculdade de Ci\^{e}ncias,
    Universidade de Lisboa,
    Av.\ Prof.\ Gama Pinto 2,
    1649-003 Lisboa, Portugal}
\author{P.\ M.\ Ferreira}\thanks{E-mail: ferreira@cii.fc.ul.pt}
\affiliation{Instituto Superior de Engenharia de Lisboa,
	1959-007 Lisboa, Portugal}
\affiliation{Centro de F\'{\i}sica Te\'{o}rica e Computacional,
    Faculdade de Ci\^{e}ncias,
    Universidade de Lisboa,
    Av.\ Prof.\ Gama Pinto 2,
    1649-003 Lisboa, Portugal}
\author{Rui Santos}\thanks{E-mail: rsantos@cii.fc.ul.pt}
\affiliation{Instituto Superior de Engenharia de Lisboa,
	1959-007 Lisboa, Portugal}
\affiliation{Centro de F\'{\i}sica Te\'{o}rica e Computacional,
    Faculdade de Ci\^{e}ncias,
    Universidade de Lisboa,
    Av.\ Prof.\ Gama Pinto 2,
    1649-003 Lisboa, Portugal}
\author{Jo\~{a}o P.\ Silva}\thanks{E-mail: jpsilva@cftp.ist.utl.pt}
\affiliation{Instituto Superior de Engenharia de Lisboa,
	1959-007 Lisboa, Portugal}
\affiliation{Centro de F\'{\i}sica Te\'{o}rica de Part\'{\i}culas (CFTP),
    Instituto Superior T\'{e}cnico, Universidade T\'{e}cnica de Lisboa,
    1049-001 Lisboa, Portugal}

\date{\today}

\begin{abstract}
LHC has found hints for a Higgs particle of 125 GeV.
We investigate the possibility that such a particle is a mixture
of scalar and pseudoscalar states.
For definiteness,
we concentrate on a two Higgs doublet model with
explicit CP violation and soft $Z_2$ violation.
Including all Higgs production mechanisms,
we determine the current constraints obtained
by comparing $h \rightarrow \gamma \gamma$
with $h \rightarrow V V^\ast$,
and comment on the information which can be gained
by measurements of $h \rightarrow b \bar{b}$.
We find bounds $|s_2| \lesssim 0.83$ at one sigma,
where $|s_2|=0$ ($|s_2|=1$) corresponds to a
pure scalar (pure pseudoscalar) state.
\end{abstract}


\maketitle

\section{Introduction}

Recently, ATLAS \cite{atlas} and CMS  \cite{cms} have reported
tantalizing signals for a 125 Higgs particle.
The exact properties of this particle will be probed in the coming years.
In particular, it could be a mixture of scalar and pseudoscalar states.
Some authors have already studied the possibility that the
125 GeV particle is a pure pseudoscalar state,
both within the two Higgs doublet model (2HDM)
\cite{burdman,Cervero:2012cx}
and in a more general context
\cite{frandsen}.
However,
as the authors point out,
this possibility is at odds with the current $h \rightarrow V V^\ast$
signal,
since a pure pseudoscalar state does not couple to $V V$
(where $V=Z,W$).
Thus,
the bounds on $h \rightarrow V V^\ast$ should allow us to constrain
the amount of the pseudoscalar component in the 125 GeV Higgs.
In this article,
we study this issue in the context of a 2HDM \cite{ourreview}
with explicit CP violation and soft-breaking of the usual
$Z_2$ symmetry.
This model has been advocated in 
Refs.~\cite{Ginzburg:2002wt,Khater:2003wq,ElKaffas:2007rq,
El Kaffas:2006nt,WahabElKaffas:2007xd,Osland:2008aw,Grzadkowski:2009iz,Arhrib:2010ju}.

We concentrate on models of type I,
where all fermions couple to the same Higgs field,
and models of type II,
where the up type quarks couple to one Higgs field,
while the down type quarks and the charged leptons couple to the other.
The constraints placed by current data on the type I and type II CP conserving models
(as well as on the lepton-specific and flipped models),
have already been studied in Ref.~\cite{Ferreira:2011aa},
assuming that the 125 GeV particle is the lightest
scalar,
in Ref.~\cite{Ferreira:2012my},
under the hypothesis that the 125 GeV particle is the heaviest
scalar,
and in Ref.~\cite{burdman},
assuming a pure pseudoscalar.
These studies assume Higgs production exclusively through gluon-gluon fusion
and assume that the remaining scalar particles have
very large masses.
In our study of scalar-pseudoscalar mixing, we remove these restrictions,
considering also inclusive production through vector boson fusion,
associated production of a scalar and a vector boson and $b\bar{b}\rightarrow H$ production.
We also improve on Refs.~\cite{Ferreira:2011aa,Ferreira:2012my,burdman}
by allowing any scalar masses and mixings consistent with experiment and with
the theoretical constraints from positivity, unitarity, perturbativity,
and the oblique radiative corrections.

In section~\ref{sec:model} we describe succinctly the 2HDM
with explicit CP violation and soft $Z_2$ violation which
we will use as a concrete example of scalar-pseudoscalar mixing,
relegating to appendix~\ref{app:rates}
the formulae we have used.
In section~\ref{sec:results}
we show our main results,
and we conclude in section~\ref{sec:conclusions}.

\section{\label{sec:model}A specific model for scalar-pseudoscalar mixing}

As a specific example of scalar-pseudoscalar mixing,
we will study a model with two Higgs doublets ($\phi_1$, $\phi_2$),
with explicit CP violation,
and with soft-violation
of the $Z_2$ symmetry $\phi_1 \rightarrow \phi_1$,
$\phi_2 \rightarrow - \phi_2$.
As far as we know,
this model was first written by
Ginzburg, Krawczyk and Osland \cite{Ginzburg:2002wt}.
It was later studied in detail in Refs.~\cite{Ginzburg:2002wt,Khater:2003wq,ElKaffas:2007rq,
El Kaffas:2006nt,WahabElKaffas:2007xd,Osland:2008aw,Grzadkowski:2009iz}.
Here,
we will follow the notation of Arhrib \textit{et.\/ al} \cite{Arhrib:2010ju},
which has a very clear presentation of this model.
For ease of reference, we collect here some of its most important characteristics.

The Higgs potential of this model is
\ba
V_H
&=&
- \frac{m_{11}^2}{2} |\phi_1|^2
- \frac{m_{22}^2}{2} |\phi_2|^2
- \frac{m_{12}^2}{2}\, \phi_1^\dagger \phi_2
- \frac{(m_{12}^2)^\ast}{2}\, \phi_2^\dagger \phi_1
\nonumber\\
&&
+ \frac{\lambda_1}{2} |\phi_1|^4
+ \frac{\lambda_2}{2} |\phi_2|^4
+ \lambda_3 |\phi_1|^2 |\phi_2|^2
+ \lambda_4\, (\phi_1^\dagger \phi_2)\, (\phi_2^\dagger \phi_1)
\nonumber\\
&&
+ \frac{\lambda_5}{2} (\phi_1^\dagger \phi_2)^2
+ \frac{\lambda_5^\ast}{2} (\phi_2^\dagger \phi_1)^2,
\label{VH}
\ea
where hermiticity forces all couplings to be real, except $m_{12}^2$ and $\lambda_5$.
If the latter are complex and
$\textrm{arg}(\lambda_5) \neq 2 \textrm{arg}(m_{12}^2)$,
then there is explicit CP violation in the Higgs
potential.
In addition, we also allow for explicit CP violation in the Yukawa terms,
leading to CP violation through the CKM matrix,
as is needed to account for the current data on CP violation
in the $K$ and $B$ systems.
The $m_{12}^2$ terms constitute a soft violation of the
$Z_2$ symmetry,
which does not affect the renormalizability of the theory
(the renormalization group equations of the quartic terms do not depend on the
quadratic terms).
Naturally, one can change the phases of the $\phi_1$ and $\phi_2$ fields;
any physical observable must be rephasing invariant \cite{BLS}.
An overall phase corresponds to a global hypercharge transformation,
and has no effect on the lagrangian;
it can be used to render the vacuum expectation
value (vev) of $\phi_1$ real.
Rephasing $\phi_2$ can now be used to render its vev also real,
\be
\langle \phi_1 \rangle = v_1/\sqrt{2},
\hspace{5ex}
\langle \phi_2 \rangle = v_2/\sqrt{2},
\label{v1v2}
\ee
substantially simplifying the minimization conditions,
which become
\ba
m_{11}^2 &=&
- \textrm{Re}\left(m_{12}^2 \right) \frac{v_2}{v_1}
+ \lambda_1\, v_1^2 + \lambda_{345}\, v_2^2
\nonumber\\
m_{22}^2 &=&
- \textrm{Re}\left(m_{12}^2 \right) \frac{v_1}{v_2}
+ \lambda_2\, v_2^2 + \lambda_{345}\, v_1^2
\nonumber\\
\textrm{Im}\left(m_{12}^2 \right) &=& v_1 v_2\, \textrm{Im}\left(\lambda_5 \right),
\label{stat_cond}
\ea
where $\lambda_{345}= \lambda_3 +  \lambda_4 + \textrm{Re}\left(\lambda_5 \right)$.
With our conventions,
$v = \sqrt{v_1^2 + v_2^2} = (\sqrt{2} G_\mu)^{-1/2} = 246$ GeV.
Thus,
$v_1$ and $v_2$ depend only on $\tan{\beta} = v_2/v_1$.

We denote by ``$Z_2$ basis'',
the basis where the Higgs potential has the form in
Eq.~\eqref{VH} and the vevs are given by Eq.~\eqref{v1v2},
and we parametrize the fields in this basis by
\be
\phi_1 =
\left(
\begin{array}{c}
\varphi_1^+\\
\tfrac{1}{\sqrt{2}} (v_1 + \eta_1 + i \chi_1)
\end{array}
\right),
\hspace{5ex}
\phi_2 =
\left(
\begin{array}{c}
\varphi_2^+\\
\tfrac{1}{\sqrt{2}} (v_2 + \eta_2 + i \chi_2)
\end{array}
\right).
\ee
The charged fields are changed into the mass basis by
\ba
G^+ &=& \cos{\beta}\, \phi_1^+ + \sin{\beta}\, \varphi_2^+,
\nonumber\\
H^+ &=& - \sin{\beta}\, \phi_1^+ + \cos{\beta}\, \varphi_2^+.
\ea
We apply the same transformation to the imaginary parts of the neutral fields,
\ba
G^0 &=& \cos{\beta}\, \chi_1 + \sin{\beta}\, \chi_2,
\nonumber\\
\eta_3 &=& - \sin{\beta}\, \chi_1 + \cos{\beta}\, \chi_2,
\label{G0_eta3}
\ea
but \textit{not} to the real parts of the neutral
fields \cite{Higgs,LS}.
As shown below,
this is done in order to keep a clean definition for the angles $\alpha_i$
leading the neutral fields from the $Z_2$ basis directly into their
mass basis\footnote{Recall that $\beta$ is the angle leading from the
$Z_2$ basis into the Higgs basis. In the completely $Z_2$ symmetric case,
$\alpha - \beta$ is the angle leading from the Higgs basis into the mass basis,
and (thus) $\alpha$  is the angle leading directly from the $Z_2$ basis into the mass basis.}.
$G^+$ and $G^0$ are the would-be Goldstone bosons and $H^+$ is already
the physical charged Higgs field, with mass $m_{H^\pm}$.
Finally,
one needs to diagonalize the (squared) mass matrix
for the neutral fields ${\cal M}^2$,
whose components are
\be
\left( {\cal M}^2 \right)_{ij} =
\frac{\partial^2 V_H}{\partial \eta_i\, \partial \eta_j}.
\ee
This is achieved through an orthogonal transformation
\be
\left(
\begin{array}{c}
h_1\\
h_2\\
h_3
\end{array}
\right)
= R
\left(
\begin{array}{c}
\eta_1\\
\eta_2\\
\eta_3
\end{array}
\right),
\label{h_as_eta}
\ee
such that
\be
R\, {\cal M}^2\, R^T = \textrm{diag} \left(m_1^2, m_2^2, m_3^2 \right),
\ee
and $m_1 \leq m_2 \leq m_3$ are the masses of the neutral Higgs particles.
The matrix $R$ may be parametrized by \cite{ElKaffas:2007rq}
\be
R =
\left(
\begin{array}{ccc}
c_1 c_2 & s_1 c_2 & s_2\\
-(c_1 s_2 s_3 + s_1 c_3) & c_1 c_3 - s_1 s_2 s_3  & c_2 s_3\\
- c_1 s_2 c_3 + s_1 s_3 & -(c_1 s_3 + s_1 s_2 c_3) & c_2 c_3
\end{array}
\right)
\ee
with $s_i = \sin{\alpha_i}$ and
$c_i = \cos{\alpha_i}$ ($i = 1, 2, 3$).
Without loss of generality,
the angles may be varied in the intervals \cite{ElKaffas:2007rq}
\be
- \pi/2 < \alpha_1 \leq \pi/2,
\hspace{5ex}
- \pi/2 < \alpha_2 \leq \pi/2,
\hspace{5ex}
0 \leq \alpha_3 \leq \pi/2.
\label{range_alpha}
\ee

The lightest neutral Higgs particle
(the putative $125$ GeV state)
is determined by the first row of $R$.
If $|s_2|=0$,
then $\eta_3$ does not contribute to $h_1$,
which is a pure scalar.
Notice that,
in this case, there may be CP violation
due to mixing in the $h_2, h_3$ states \cite{J1}.
This possibility will affect CP violating observables but
not the current known data.
If $|s_2|=1$,
then only $\eta_3$ contributes to $h_1$,
which is a pure pseudoscalar.
In this case there is CP conservation.
We conclude that
\ba
|s_2| = 0
& \Rightarrow &
h_1\ \textrm{is a pure scalar},
\label{pure_scalar}
\\
|s_2| = 1
& \Rightarrow &
h_1\ \textrm{is a pure pseudoscalar},
\label{pure_pseudoscalar}
\ea
and $|s_2|$ is a measure of the pseudoscalar content of the
lightest Higgs scalar.

Given Eqs.~\eqref{stat_cond},
the scalar sector depends only on
the eight parameters
$\beta$,
$\textrm{Re}(m_{12}^2)$,
$\textrm{Im}(m_{12}^2)$,
$\lambda_{1,2,3,4}$,
and $\textrm{Re}(\lambda_5)$.
As suggested in Ref.~\cite{El Kaffas:2006nt},
these can be traded for $m_1$, $m_2$,
$m_{H^\pm}$, $\alpha_{1,2,3}$, $\beta$,
and $\textrm{Re}(m_{12}^2)$.
In this approach,
$m_3$ is a derived quantity,
given by
\be
m_3^2 = \frac{m_1^2\, R_{13} (R_{12} \tan{\beta} - R_{11})
+ m_2^2\ R_{23} (R_{22} \tan{\beta} - R_{21})}{R_{33} (R_{31} - R_{32} \tan{\beta})}.
\label{m3_derived}
\ee

The implementation of the $Z_2$ symmetry in the fermion sector
means that each fermion type (up quark, down quark, and charged leptons)
can couple only to one of the original scalar doublets.
As an example,
let us consider three down type quarks coupling exclusively to $\phi_1$.
We start from the Yukawa lagrangian
\be
- {\cal L}_Y = \bar{q}_L \Gamma_1 \phi_1 n_R + \textrm{h.c.},
\label{Ly1}
\ee
where $q_L^T = (p_L, n_L) $ is a left-handed doublet of the gauge group,
having $3$ components in family space,
while $n_R$ has $3$ down type quarks,
each a singlet under the gauge group.
$\Gamma_1$ is the $3 \times 3$ matrix of Yukawa couplings,
and $\textrm{h.c.}$ stands for hermitian conjugation.
After SSB,
the lagrangian has a piece involving the neutral component of $\phi_1$,
which may be written as
\be
\bar{n}_L \Gamma_1 \frac{v_1}{\sqrt{2}}
\left[ 1 + \frac{\eta_1 + i \chi_1}{v_1} \right] n_R + \textrm{h.c.}
=
\bar{d}_L M_d d_R + \bar{d}_L \frac{M_d}{v_1}  \left( \eta_1 + i \chi_1 \right) d_R + \textrm{h.c.}.
\label{Ly2}
\ee
To obtain the last expression we have rotated the fields $n_L$ and $n_R$
into the mass basis ($d_L$ and $d_R$),
diagonalizing the matrix
$\Gamma_1 v_1 /\sqrt{2}$ to obtain the (diagonal) down quark mass matrix $M_d$.
Now,
we need
\ba
\eta_1 \pm i \chi_1
&=&
\left( R_{k1} \mp i s_\beta R_{k3} \right) h_k \pm i c_\beta G^0,
\nonumber\\
\eta_2 \pm i \chi_2
&=&
\left( R_{k2} \pm i c_\beta R_{k3} \right) h_k \pm i s_\beta G^0,
\label{comb}
\ea
which were obtained by inverting Eqs.~\eqref{G0_eta3} and \eqref{h_as_eta},
and where a sum over $k = 1,2,3$ is implied.
The couplings with the neutral scalar particles are obtained substituting
Eq.~\eqref{comb} in Eq.~\eqref{Ly2} to get
\be
\bar{d}_L \frac{M_d}{v_1}
\left(
R_{k1} - i s_\beta R_{k3}
\right)
h_k d_R + \textrm{h.c.}
=
\bar{d}_L \frac{M_d}{v}
\left(
\frac{R_{k1}}{c_\beta} - i s_\beta \frac{R_{k3}}{c_\beta} \gamma_5
\right)
d_R h_k.
\label{Ly3}
\ee
We are interested in the couplings of the fermions to the lightest scalar $h_1$.
Comparing with the notation of the effective lagrangian \eqref{LY}
in the appendix, needed for our calculations,
we obtain
\be
a = \frac{R_{11}}{c_\beta},
\hspace{5ex}
b = - s_\beta \frac{R_{13}}{c_\beta}
\hspace{5ex}
\left( \textrm{down type quarks couple to } \phi_1 \right).
\label{down1}
\ee
A similar analysis leads to Table~\ref{tab:couplings}.
\begin{table}
\begin{center}
\begin{tabular}{ccccccccc}
\hline
 & & Type I  & & Type II & & Lepton & & Flipped \\
 & & & & & & Specific & & \\
\hline
Up  & &
$\tfrac{R_{12}}{s_{\beta}} - i c_\beta \tfrac{R_{13}}{s_{\beta}}$    & &
$\tfrac{R_{12}}{s_{\beta}} - i c_\beta \tfrac{R_{13}}{s_{\beta}}$    & &
$\tfrac{R_{12}}{s_{\beta}} - i c_\beta \tfrac{R_{13}}{s_{\beta}}$    & &
$\tfrac{R_{12}}{s_{\beta}} - i c_\beta \tfrac{R_{13}}{s_{\beta}}$    \\
Down  & &
$\tfrac{R_{12}}{s_{\beta}} + i c_\beta \tfrac{R_{13}}{s_{\beta}}$    & &
$\tfrac{R_{11}}{c_{\beta}} - i s_\beta \tfrac{R_{13}}{c_{\beta}}$     & &
$\tfrac{R_{12}}{s_{\beta}} + i c_\beta \tfrac{R_{13}}{s_{\beta}}$    & &
$\tfrac{R_{11}}{c_{\beta}} - i s_\beta \tfrac{R_{13}}{c_{\beta}}$     \\
Leptons  & &
$\tfrac{R_{12}}{s_{\beta}} + i c_\beta \tfrac{R_{13}}{s_{\beta}}$    & &
$\tfrac{R_{11}}{c_{\beta}} - i s_\beta \tfrac{R_{13}}{c_{\beta}}$     & &
$\tfrac{R_{11}}{c_{\beta}} - i s_\beta \tfrac{R_{13}}{c_{\beta}}$    & &
$\tfrac{R_{12}}{s_{\beta}} + i c_\beta \tfrac{R_{13}}{s_{\beta}}$    \\
\hline
\end{tabular}
\end{center}
\caption{Couplings of the fermions to the lightest scalar, $h_1$,
presented, for each case, in the form $a + i b$.
\label{tab:couplings}}
\end{table}
Notice that $s_2=0$ implies $R_{13}=0$,
in which case all $b$ coefficients in Table~\ref{tab:couplings} vanish,
confirming that $h_1$ couples to the fermions as a pure scalar,
and vindicating Eq.~\eqref{pure_scalar}.
Similarly,
$|s_2|=1$ leads to $R_{11}=0=R_{12}$,
and all $a$ coefficients in Table~\ref{tab:couplings} vanish,
implying that $h_1$ couples to the fermions with $i \gamma_5$,
being a pure pseudoscalar, as stated in Eq.~\eqref{pure_pseudoscalar}.

In Model I, all quarks and charged leptons couple to the doublet $\phi_2$,
and the corresponding coupling factors can be read from the first column
of Table~\ref{tab:couplings}. In Model II,
the up quarks still couple to $\phi_2$ but the remaining fermions
couple to $\phi_1$ - the respective couplings are shown in the
second column of Table~\ref{tab:couplings}.
The Lepton-Specific and Flipped models discussed in
Refs.~\cite{Ferreira:2011aa,Ferreira:2012my}
would be obtained by taking for the charged leptons the opposite
choice taken for the down quarks.
We have checked that our results reproduce those for Model II
included in Ref.~\cite{Arhrib:2010ju}.

Expanding the covariant derivative terms of the neutral scalars,
the triple interactions of $h_1$ with $WW$ and $ZZ$ may be written
as in Eq.~\eqref{LVV},
with
\be
C = c_\beta R_{11} + s_\beta R_{12}.
\label{C_formula}
\ee
As expected,
when $|s_2|=1$,
$R_{11}=0=R_{12}$ and there is no tree-level coupling of the
pure pseudoscalar $h_1$ to a pair of gauge bosons.
From the Higgs potential, we may get the triple vertex of $h_1$ with
the charged Higgs bosons as in Eq.~\eqref{LhHpHm},
with
\be
- \lambda
=
c_\beta \left[ s_\beta^2 \lambda_{145} + c_\beta^2 \lambda_3 \right] R_{11}
+
s_\beta \left[ c_\beta^2 \lambda_{245} + s_\beta^2 \lambda_3 \right] R_{12}
+
s_\beta c_\beta\, \textrm{Im}(\lambda_5)\, R_{13},
\label{lambda_hHpHm}
\ee
where $\lambda_{145} = \lambda_1 - \lambda_4 - \textrm{Re}(\lambda_5)$
and $\lambda_{245} = \lambda_2 - \lambda_4 - \textrm{Re}(\lambda_5)$.
In the pure pseudoscalar limit,
only the last term of Eq.~\eqref{lambda_hHpHm} survives,
showing that a pure pseudoscalar can only couple to a pair
of charged scalars if there is explicit CP violation
in the scalar potential, through $\textrm{Im}(\lambda_5)$.

For a given set of input parameters,
the effective couplings ($a$, $b$, $C$, and $\lambda$) discussed in this section
can be used on the equations in appendix~\ref{app:rates} in order to
find the production rates for $h_1$ and its decay rates into all final states.
For each final state $f$, we define the ratio
\be
R_f = \frac{\sigma (p p \rightarrow h_1)\ \textrm{BR} (h_1 \rightarrow f)}{
\sigma (p p \rightarrow h)_{\textrm{SM}}\ \textrm{BR} (h \rightarrow f)_{\textrm{SM}}},
\ee
where $\sigma (p p \rightarrow h_1)$ is an inclusive production rate obtained by summing
over
production mechanisms of $h_1$,
and $\textrm{BR} (h_1 \rightarrow f)$
is the branching ratio of the decay of $h_1$ into the final state $f$.

The production mechanisms included in $\sigma (p p \rightarrow h_1)$ include:
the usual gluon-gluon fusion processes (which involve loops with both
top and bottom quarks); vector boson fusion (VBF) processes; so-called associated production
processes, with a $W$ or $Z$ boson in the final state, alongside $h_1$;
and the $b\bar{b} \rightarrow h_1$ process as well. An important point
needs to be explained at this juncture: since $h_1$ is a mix of scalar
and pseudoscalar states, its production mechanism will involve, likewise,
a mix of production rates pertaining to a scalar and a pseudoscalar
particle. For instance, a pure pseudoscalar particle has no VBF or associated
production mechanisms; on the other hand, due to different fermion couplings,
the gluon-gluon fusion cross section is different for production of
a scalar or pseudoscalar states of the same mass. The exact formulae for
$\sigma (p p \rightarrow h_1)$ can be found in Appendix~\ref{app:rates}.
Using those formulae we were able to express the production factors in terms 
of next-to-leading order cross sections for each of the mechanisms
considered, calculated using HIGLU~\cite{Spira:1995mt} for the gluon-gluon
cross section, bb@nnlo~\cite{Harlander:2003ai} for the $b \bar b$ process
and reference~\cite{Baglio:2010ae} (and references therein), for the remaining processes,
all multiplied by the adequate factors
pertaining to the mixing of scalar-pseudoscalar states.

The branching ratios $\textrm{BR} (h_1 \rightarrow f)$ also need to
be computed considering the mixed nature of the $h_1$ state, and the
relevant formulae can be found in Appendix~\ref{app:rates}. Notice
that, since the branching ratio is the partial width divided by
the sum of all decay widths, a parameter choice that affects,
for example,
$h_1 \rightarrow b \bar{b}$ will have an impact on
$R_{\gamma \gamma}$ because it affects the overall decay width.

We are particularly interested on what one can learn
from the current LHC bounds on $R_{\gamma \gamma}$,
$R_{ZZ}$, $R_{WW}$, and $R_{b \bar{b}}$.
These are combined with other known constraints,
including recent results from the Tevatron \cite{tevatron,TEVNPH:2012ab}.
In this analysis,
we utilize the explicit bounds obtained for the several $R_f$
from the experimental data by
Espinosa \textit{et. al} \cite{Espinosa:2012ir}. Specifically,
we use their Table III, updated after Moriond 2012.
Similar summaries of the experimental bounds
have been obtained in Refs.~\cite{Giardino:2012ww,Carmi:2012yp,Azatov:2012bz}.
Since the current errors are large,
we are not interested so much in the precise values but rather in
the qualitative features hinted at by current bounds.
Surprisingly,
we can already exclude very large regions of the 2HDM parameter
space, and place constraints on the pseudoscalar content
of the lightest Higgs particle.

\section{\label{sec:results}Analysis and results}

We performed an extensive scan of the 8-dimensional parameter space of the
models we are studying.
Our fixed inputs are $v = 246$ GeV and $m_1 = 125$ GeV.
We then took random values of:
the angles $\alpha_1$, $\alpha_2$ and $\alpha_3$ in their allowed intervals
of variation, specified in Eq.~\eqref{range_alpha};
$\tan\beta$ between 1 and
30;
the masses $m_2$ (above the value of $m_1$) and
$m_{H^\pm}$, the latter with values above 90 GeV;
and $\textrm{Re}(m^2_{12})$, taken in the range from  $-10^6$ to $+10^6$ GeV$^2$.

The charged Higgs couplings to the fermions have exactly the same form as the ones
in a softly broken $Z_2$ symmetric two-Higgs doublet model. Therefore the bounds
derived for the charged sector still hold for the CP-violating scenario under
study.
The LEP $R_b$ constraint (from $Z \rightarrow b\bar{b}$) 
excludes values of $\tan \beta < 1$ even in the CP violating
scenario. One should note that although the expressions
for $R_b$ change in the CP-violating scenario, the main contributions for low
$\tan \beta$ come from the charged Higgs diagrams~\cite{ElKaffas:2008zz,WahabElKaffas:2007xd}.
Constraints from B-physics,
and particularly those coming from
$b\rightarrow s \gamma$~\cite{Hou:1987kf,Ciuchini:1997xe,Borzumati:1998tg,
Borzumati:1998nx,Amsler:2008zzb,Limosani:2009qg}, have excluded a
charged Higgs boson mass  below 300 GeV in models type II and Flipped, almost
independently of
$\tan \beta$. Charged Higgs bosons with masses as low as 100 GeV are
instead still allowed in models Type I and
Lepton-Specific~\cite{Aoki:2009ha,Logan:2009uf,Su:2009fz}. Finally
$b\rightarrow s \gamma$ implies $\tan \beta > 1$ for a charged Higgs mass
below approximately 600 GeV.

Once a given set of parameters is chosen, a value for $m_3^2$ is computed from
Eq.~\eqref{m3_derived}.
If that value is positive and larger than $m_2^2$,
we then verify that all theoretical constraints
on the potential are satisfied.
The quartic couplings $\lambda_{1 ... 5}$
are computed using the formulae in Eq.~(B.1) from Ref.~\cite{Arhrib:2010ju}.
One then verifies whether these quartic couplings obey the conditions
which ensure that the potential is bounded from below and that unitarity and perturbativity
are satisfied - Eqs.~(2.20) and (2.21) from Ref.~\cite{Arhrib:2010ju},
respectively.
Finally, one verifies that the potential's parameters are such that the
current constraints on the $S$, $T$ and $U$ oblique parameters are obeyed.
To do so, we have used the formulae presented in Appendix D of Ref.~\cite{ourreview}.
If all of these constraints are obeyed,
the set of parameters (a ``point" in parameter space)
is deemed satisfactory and one then proceeds to calculate the
branching ratios and production factors of the $h_1$ state.

\subsection{Model I}
\label{sec:modI}

In Model I, as was explained earlier, all fermions couple to the same doublet,
$\phi_2$ by convention. The couplings between $h_1$ and the fermions are affected
by the $a$ and $b$ parameters shown in eq.~\eqref{LY} which, for this
model and the $\tan\beta$ we chose, tend to be smaller than 1. We generated
over 270000 points for this model.

\begin{figure}[htb]
\centering
\includegraphics[height=9cm,angle=0]{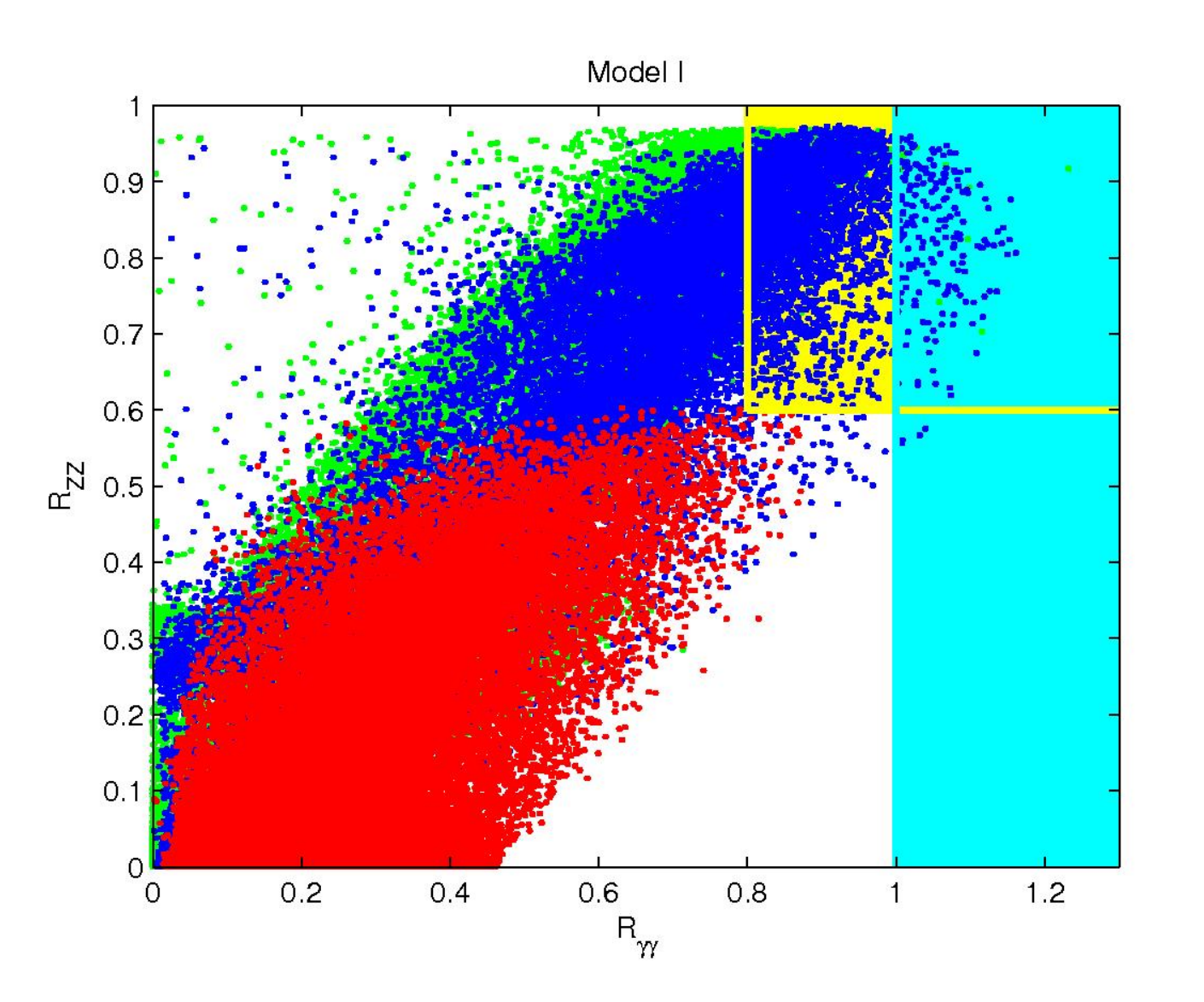}
\caption{Scatter plot in the $R_{\gamma \gamma} - R_{ZZ}$ plane
for the type I 2HDM.
The color coded points have the following correspondences:
green (light grey) means $|s_2| < 0.1$ ($h_1$ is mostly scalar),
blue (black) means $0.45 < |s_2| < 0.55$,
and red (dark grey) means $|s_2| > 0.83$ ($h_1$ is mostly pseudoscalar).
The yellow and light blue (light grey and grey) bands shows the current
ATLAS and  CMS bounds, from Ref.~\cite{Espinosa:2012ir}.
}
\label{RZZphph_modI}
\end{figure}

In Fig.~\ref{RZZphph_modI}, we show how the $R_{\gamma \gamma} - R_{ZZ}$ plane
is filled by Model I, for three selections of points: points for which one has
an $h_1$ state which is essentially a scalar ({\em i.e.}, for which the
angle $\alpha_2$ obeys $|s_2| < 0.1$); an $h_1$ state with $0.45 < |s_2| < 0.55$;
and points for which $h_1$ is essentially a pseudoscalar ($|s_2| > 0.83$).
There are several salient features.
First, we notice that even the current rather loose bounds already
kill large regions of the model's parameter space.
In particular,
$R_{ZZ}$ tends to be smaller than 1 in Model I,
for any values of the parameters of the model.
Second,
a large pseudoscalar component ($|s_2| > 0.83$) is already
excluded,
both by the $ZZ$ bounds and by the $\gamma \gamma$ bounds.
That the $ZZ$ bounds constrain $s_2$ was expected
and was the primary motivation for this work.
Indeed,
when $|s_2|=1$,
Eq.~\eqref{C_formula} leads to $C=0$,
guaranteeing that there is no coupling of $h_1$ to $ZZ$ nor to $WW$.
What is new is that,
in this model,
the known experimental bounds and
theoretical constraints 
make $|s_2|\sim 1$ inconsistent with $R_{\gamma \gamma}$
\textit{even if $R_{ZZ}$ vanished}.
A third important point
(not clearly visible in Fig.~\ref{RZZphph_modI})
is that the $0.45 < |s_2| < 0.55$ (blue/black) region extends further into the
$R_{\gamma \gamma}$ experimentally allowed region than the $|s_2| < 0.1$ (green/light grey) 
region~\footnote{Please notice that there is some superposition of points in several regions,
so that ``underneath" the points shown in blue/black, for instance, there may well exist
green/light grey points. This is common to all figures shown in this work.}.
This has the following implication:
if the current ATLAS central values for $R_{\gamma \gamma}$
remain as the errors get smaller,
then the model seems to prefer a mixture of scalar and pseudoscalar
components in the lowest lying Higgs particle.
Finally, we find that only for $|s_2| < 0.83$
do we start to get points inside the band
allowed by current experiments.

\begin{figure}[htb]
\centering
\hspace{-1.cm}
\includegraphics[height=9cm,angle=0]{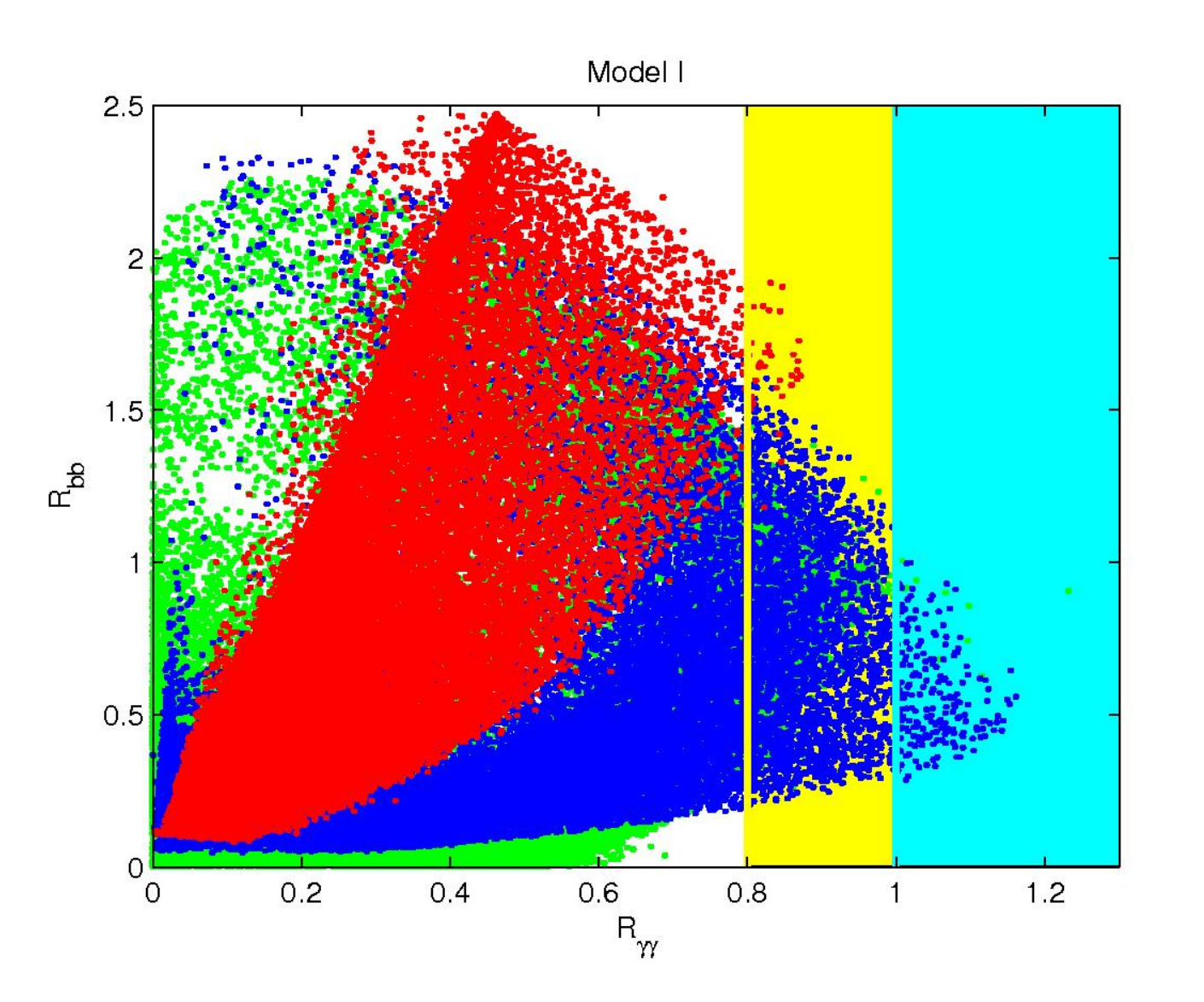}
\vspace{-0.4cm}
\caption{Scatter plot in the $R_{\gamma \gamma} - R_{b \bar{b}}$ plane
for the type I 2HDM.
The color codes are the same as the previous figure's.
}
\label{Rbbphph_modI}
\end{figure}

We have mentioned that
there are very few points above around $R_{\gamma \gamma} = 1$
corresponding to the (almost) pure scalar solution
$|s_2| < 0.1$.
There are quite a few more points in the large $R_{\gamma \gamma}$
region corresponding to $0.45 < |s_2| < 0.55$.
Fig.~\ref{RZZphph_modI} shows that the former tend to concentrate around
$R_{ZZ} \sim 0.8$.
A better handle on the difference between
$|s_2| < 0.1$ and  $0.45 < |s_2| < 0.55$ is provided by
$R_{b \bar{b}}$,
as shown in Fig.~\ref{Rbbphph_modI},
where Model I's results are presented
in the $R_{\gamma \gamma} - R_{b \bar{b}}$ plane.
We see that the pure scalar solution with large $R_{\gamma \gamma}$
leads to $R_{b \bar{b}} \sim 1$,
while the large values of $R_{\gamma \gamma}$
correspond to $R_{b \bar{b}} \sim 0.5$ when
$0.45 < |s_2| < 0.55$.
Thus, as the errors get smaller,
a comparison between $R_{\gamma \gamma}$, $R_{ZZ}$,
and $R_{b \bar{b}}$ can be used to further constrain
$|s_2|$ in Model I. 

Regarding the current bounds on $R_{b \bar{b}}$, Espinosa
{\em et al}~\cite{Espinosa:2012ir} provide an LHC
interval of $R_{b \bar{b}} < 3.3$, which spans
the entire range for this variable displayed in
Fig.~\ref{Rbbphph_modI}. In that reference there is also a bound
stemming from Tevatron data, $1.3 < R^{TEVATRON}_{b \bar{b}} < 2.8$;
once translated into LHC bounds for Model I - that is, with the appropriate
LHC production factors - this would give approximately
$R_{b \bar{b}} > 1.2$ in Fig.~\ref{Rbbphph_modI}. This would
eliminate most of the available parameter space left by the
bounds on $R_{\gamma \gamma}$. Accurate measurements of
$R_{b \bar{b}}$ are thus of extreme importance for
this model.

We have also looked at $h_1 \rightarrow W^+ W^-$.
Notice that, within our tree-level calculations, and even for
a mixed $h_1$ state, $R_{WW} \simeq R_{ZZ}$, since we are
dealing with ratios to SM quantities. As such,
one can glean information about $R_{WW}$ from the plot in
Fig.~\ref{RZZphph_modI}.
After an earlier central value compatible with the SM,
ATLAS presented at Moriond 2012 results which are
now slightly over one sigma smaller than the SM
-- compare the respective entries in Tables I and III
of Ref.~\cite{ Espinosa:2012ir}.
Their Moriond 2012 bounds impose $R_{WW}$ smaller than
0.8 for ATLAS and 1 for CMS.
If this situation were to remain as errors get smaller,
then the SM itself would be in trouble.
As seen in Fig.~\ref{RZZphph_modI} for $h_1 \rightarrow Z Z$
in Model I,
the values for $ h_1 \rightarrow  W^+ W^-$ are smaller than one,
enabling a possible accommodation of a low $R_{WW}$ signal.

\subsection{Model II}
\label{sec:modII}

We now turn to Model II. As explained, in this model
the down-type quarks and charged leptons couple to the $\phi_1$ doublet,
and the up-type quarks couple to $\phi_2$. The corresponding couplings
can be read off Table~\ref{tab:couplings}, and we see that the bottom
quark couplings are proportional to $1/\cos\beta$. As such, one can expect,
in this model, an enhancement of the production of $h_1$ via
the gluon-gluon fusion $b$-quark triangle, or the $b\bar{b} \rightarrow h_1$
channel for large values of $\tan\beta$. Similarly, the branching ratio
$BR(h_1 \rightarrow b\bar{b})$ will also reach higher values than in Model I
for large $\tan\beta$. As was mentioned earlier, in Model II there are
stringent constraints on the value of the charged Higgs mass - it needs to
be larger than about 300 GeV - stemming from $b\rightarrow s \gamma$
data. We included that cut in our scan of this model's parameter space,
and again we generated
over 270000 points for this model.

\begin{figure}[htb]
\centering
\hspace{-1.cm}
\includegraphics[height=9cm,angle=0]{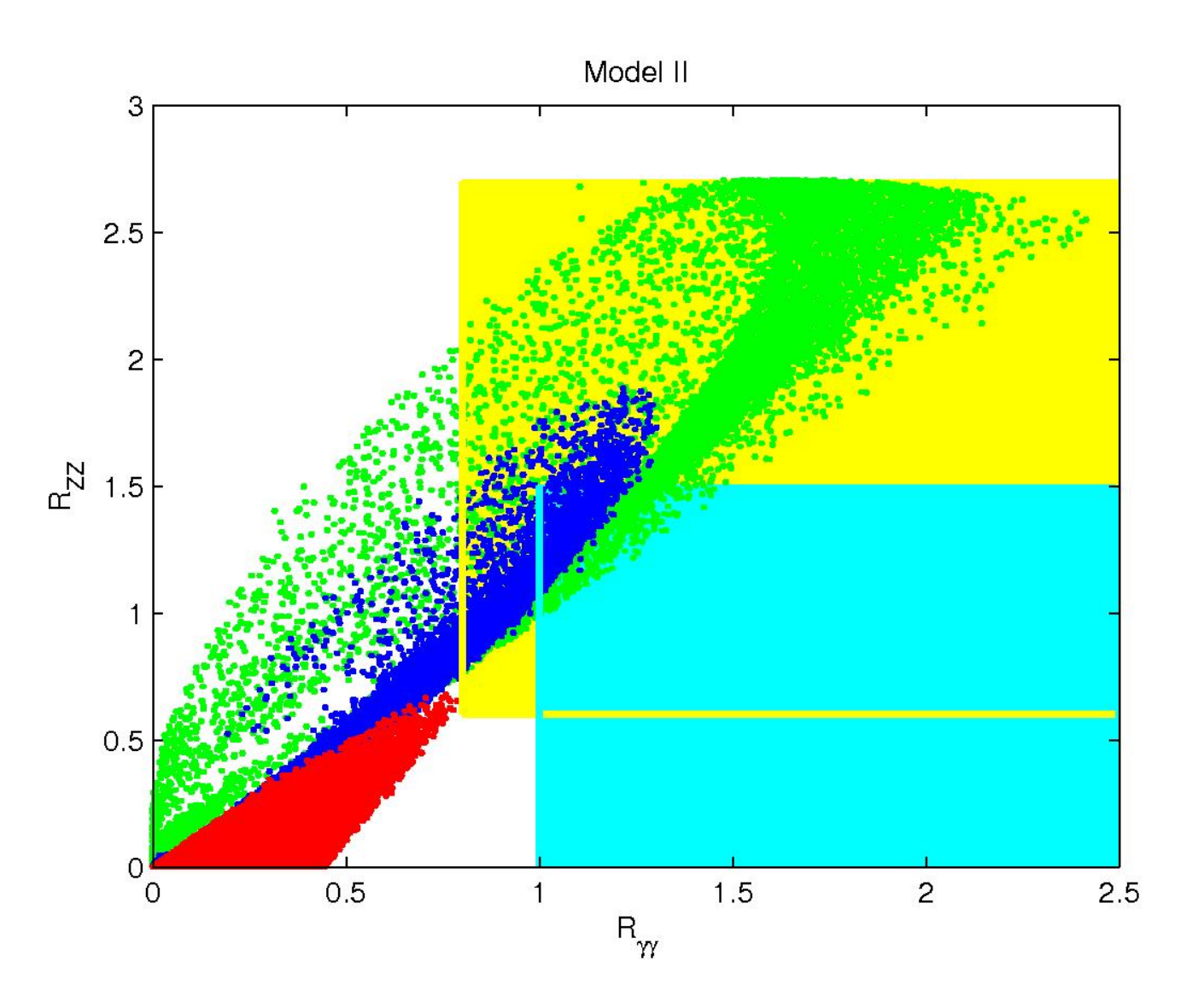}
\vspace{-0.4cm}
\caption{Scatter plot in the $R_{\gamma \gamma} - R_{ZZ}$ plane
for the type II 2HDM.
The color codes are the same as previous figures'.
}
\label{RZZphph_modII}
\end{figure}

\begin{figure}[htb]
\centering
\hspace{-1.cm}
\includegraphics[height=9cm,angle=0]{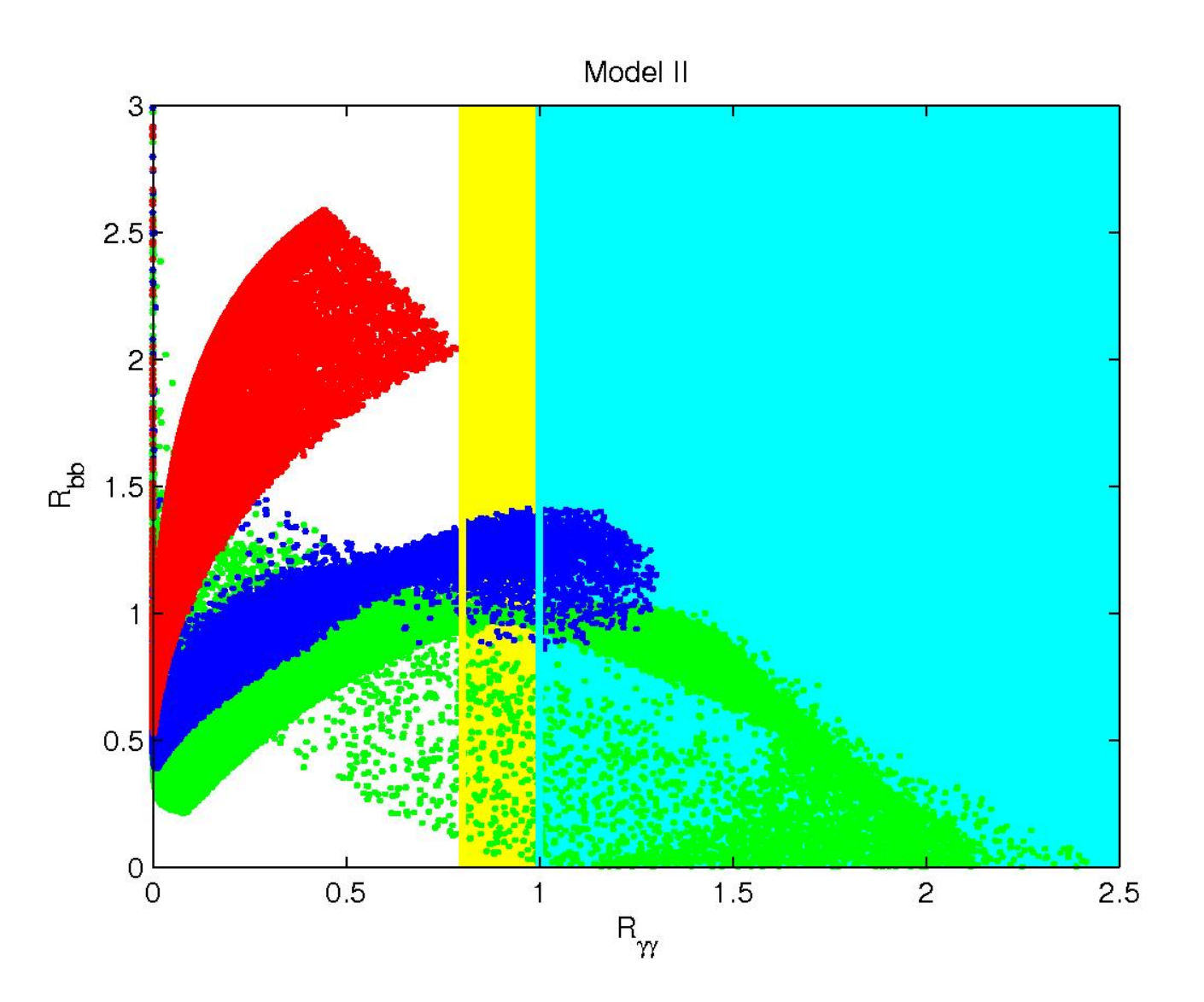}
\vspace{-0.4cm}
\caption{Scatter plot in the $R_{\gamma \gamma} - R_{b \bar{b}}$ plane
for the type II 2HDM.
The color codes are the same as previous figures'.
}
\label{Rbbphph_modII}
\end{figure}

In Fig.~\ref{RZZphph_modII} we show our results for Model II in the
$R_{\gamma \gamma} - R_{ZZ}$ plane,
while
Fig.~\ref{Rbbphph_modII} shows the results in the
$R_{\gamma \gamma} - R_{b \bar{b}}$ plane.
As expected,
large values of $|s_2|$ are excluded by the $ZZ$ bound.
There are similarities and differences between the two models.
The striking feature that $R_{\gamma \gamma}$ by itself constrains
$|s_2|$ is common to models I and II,
and in both cases $|s_2|$ is found to have to be smaller than
about 0.83.

The most noticeable difference between both models is that
Model I keeps roughly
$R_{\gamma\gamma} < 2$,
$R_{ZZ} < 1$,
and $R_{b \bar{b}} < 2.5$
while, in Model II,
values as large as
$R_{\gamma\gamma} < 2.5$,
$R_{ZZ} < 2.7$,
and $R_{b \bar{b}} > 2.5$~\footnote{Values as high as 10
for $R_{b \bar{b}}$ were found, for extremely small values of
$R_{\gamma\gamma}$, though they are not displayed in the plot of
Fig.~\ref{Rbbphph_modII} for ease of presentation. These high
values correspond to both the gauge-phobic limit of this
model and the enhancement with large $\tan\beta$ alluded to earlier.}
are allowed.
Also,
in Model II,
the $|s_2| < 0.1$ (green/light grey) region extends further into the
$R_{\gamma \gamma}$ experimentally allowed region than
the $0.45 < |s_2| < 0.55$ (blue/black) region.
This is the opposite of what we observe in Model I.
Finally,
we see from Fig.~\ref{Rbbphph_modII}
that the pure scalar solution with large $R_{\gamma \gamma}$
tends to imply $R_{b \bar{b}} < 1$,
while the large values of $R_{\gamma \gamma}$
correspond to $R_{b \bar{b}} > 1$ when
$0.45 < |s_2| < 0.55$.
Thus, as better measurements are available,
$R_{b \bar{b}}$ might become instrumental
in constraining $s_2$. In fact, the current Tevatron bounds,
cited by~\cite{Espinosa:2012ir}, translated into LHC
for Model II, would give roughly $1.1 < R_{b \bar{b}} < 2.8$,
which, if confirmed, would exclude an extremely significative
portion of this model's parameter space. In particular, it would
seem to disfavor the possibility of $h_1$ being a pure scalar.

Finally, a word on the constraints emerging from $R_{WW}$ for this model.
As mentioned above for the Model I analysis, current ATLAS and CMS bounds
favor values of $R_{WW}$ smaller than 1 - since $R_{WW} \simeq R_{ZZ}$,
we can see from Fig.~\ref{RZZphph_modII} that that region is
heavily disfavoured for Model II
by the current bounds on $R_{\gamma\gamma}$.
As a result,
very low experimental values for $R_{WW}$
will exclude the SM and also the type II model.

\section{\label{sec:conclusions}Conclusions}

With the LHC providing physicists with a wealth of data, and the first hints
of the existence of a Higgs particle with a mass around 125 GeV, it becomes
possible to constrain, not only the SM but also extensions of it, such as the
2HDM. Previous studies considered the possibility that the LHC might be observing a pure
scalar, or a pure pseudoscalar. We have considered the possibility that the
putative scalar candidate at the LHC is a mixed state, neither scalar nor
pseudoscalar. We considered a specific model, a version of the 2HDM with a 
$Z_2$ discrete symmetry which has been softly broken, such that the model
has explicit CP violation in the scalar sector. The degree of ``pseudoscalarity" of
the lightest scalar $h_1$ is measured by a mixing angle $\alpha_2$ such that
$|s_2| = |\sin\alpha_2| \simeq 1$ corresponds to a pure pseudoscalar state (and
$|s_2| \simeq 0$ to a pure scalar one). We considered two specific extensions of
the $Z_2$ symmetry to the fermionic sector, the so-called Models I and II,
which have very different phenomenologies.

We have computed the production rate of $h_1$ times its branching ratio 
into several final states - $R_f$ - relative to the expected values for
such observables in the SM. We have shown that this version of the 2HDM
can do at least as good a job as the SM in fitting the current data. But 
we have also shown that even the current loose bounds 
on $R_{\gamma\gamma}$ and $R_{ZZ}$ already put severe constraints on 
the parameter space of these versions of the 2HDM. 
In particular, our work suggests
that if current trends in the values of $R_{WW}$ at the LHC, as well as
in the Tevatron data on $R_{b\bar{b}}$, persist, the versions of
the 2HDM herein considered might have a hard time reproducing the data.
However, if that were the case, the SM would also be in trouble.
Current data forces $|s_2|<0.83$, a constraint which
holds roughly even if one considers only the $R_{\gamma\gamma}$ bound.
This by itself excludes a large pseudoscalar component.

\begin{acknowledgments}
We are grateful to A. Arhrib for discussions pertaining
to his related study \cite{friend}.
We thank Renato Guedes for help
with HIGLU.
J.P.S. is grateful to J.C. Rom\~{a}o for useful
discussions.
This work is supported in part by the Portuguese
\textit{Funda\c{c}\~{a}o para a Ci\^{e}ncia e a Tecnologia} (FCT)
under contract PTDC/FIS/117951/2010 and by an
FP7 Reintegration Grant, number PERG08-GA-2010-277025.
P.M.F. and R.S. are also
partially supported by PEst-OE/FIS/UI0618/2011.
The work of J.P.S. is also funded by FCT through the projects
CERN/FP/109305/2009 and  U777-Plurianual,
and by the EU RTN project Marie Curie: PITN-GA-2009-237920.
\end{acknowledgments}

\vspace{2ex}

\textbf{Note added:} While writing this paper,
Ref.~\cite{mader} appeared discussing a complementary and interesting feature
of the type I model: the production and detection of its charged Higgs.

\appendix
\section{\label{app:rates}Production and decay Rates}

In this appendix we present the decay rates for a particle $h$ with both
scalar and pseudoscalar components.
The relevant pieces of the Lagrangian are:
\ba
 {\cal L}_Y &=& -\left(\sqrt{2} G_\mu \right)^{\tfrac{1}{2}}\, m_f\
\bar \psi \left( a + i b \gamma_5 \right) \psi\, h,
\label{LY}
\\
{\cal L}_{h H^+ H^-} &=& \lambda\, v\, h\ H^+ H^-,
\label{LhHpHm}
\\
{\cal L}_{h V V} &=& C \left[
g\, m_W W_\mu^+ W^{\mu -} + \frac{g}{2 c_W} m_Z Z_\mu Z^\mu
\right]\, h,
\label{LVV}
\ea
where $a$, $b$, and $C$ are real, $c_W = \cos{\theta_W}$,
and $\theta_W$ is the Weinberg angle.
In the SM, $a=C=1$, and $b = \lambda=0$.

We find:
\ba
\Gamma(h \rightarrow \gamma \gamma)
&=&
\frac{G_\mu \alpha^2 M_h^3}{128 \sqrt{2} \pi^3}
\left\{
\left|
\sum_f N_c Q_f^2\, a\, A_{1/2}(\tau_f)
+
C A_{1}(\tau_W)
-
\frac{v^2}{2 m_{H^\pm}^2 }\, \lambda\, A_0(\tau_{\pm})
\right|^2
\right.
\nonumber\\
& &
\hspace{12ex}
+
\left.
\left|
\sum_f N_c Q_f^2\, b\, A_{1/2}^A(\tau_f)
\right|^2
\right\},
\ea
where $v= \left[\sqrt{2} G_\mu\right]^{-1/2} \approx 246\, \textrm{GeV}$
and $m_{H^\pm}$ is the mass of the charged Higgs.
For ease of reference,
we have used a notation close to that of Djouadi \cite{djouadi1, djouadi2},
where
\ba
A_{1/2}(\tau) &=& 2 [\tau + (\tau - 1) f(\tau)] \tau^{-2},
\\
A_{1/2}^A(\tau) &=& 2 \tau^{-1} f(\tau),
\\
A_1(\tau) &=& -[2 \tau^2 + 3 \tau + 3(2 \tau-1)f(\tau)] \tau^{-2}
\\
A_0(\tau) &=& - [\tau - f(\tau)] \tau^{-2},
\ea
and
\be
f(\tau)
=
\left\{
\begin{array}{ll}
\left[ \arcsin(\sqrt{\tau}) \right]^2
& \tau \leq 1\\*[2mm]
-\frac{1}{4} \left[ \log{\frac{1 + \sqrt{1-\tau^{-1}}}{1 - \sqrt{1-\tau^{-1}}}} - i \pi \right]^2
& \tau > 1
\end{array}
\right.
.
\ee
The scaling variables are $\tau_i = M_h^2/(4 m_i^2)$,
where $m_i$ is the mass of the particle in the loop.

Similarly,
for the decays into two gluons we find
\ba
\Gamma(h \rightarrow g g)
&=&
\frac{G_\mu \alpha_s^2 M_h^3}{64 \sqrt{2} \pi^3}
\left\{
\left|
\sum_q a\, A_{1/2}(\tau_q)
\right|^2
+
\left|
\sum_q b\, A_{1/2}^A(\tau_q)
\right|^2
\right\},
\ea
where the sums run only over quarks $q$.

The decays into fermions are given by
\be
\Gamma(h \rightarrow f \bar{f})
=
N_c \frac{G_\mu\, m_f^2}{4 \sqrt{2} \pi} M_h
\left[ a^2 \beta_f^3 + b^2 \beta_f \right],
\ee
where $\beta_f = \sqrt{1 - 4 m_f^2/M_h^2} = \sqrt{1 - \tau^{-1}}$,
while the decays into two vector bosons are given by
\be
\Gamma(h \rightarrow V^{(\ast)} V^{(\ast)})
=
C^2\
\Gamma_{\textrm{SM}}(h \rightarrow V^{(\ast)} V^{(\ast)}),
\ee
and the partial decay widths in the SM-Higgs case in the two-, three-
and four-body approximations,
$\Gamma_{\textrm{SM}}(h \rightarrow V^{(\ast)} V^{(\ast)})$,
can be found in Section I.2.2 of Ref.~\cite{djouadi1}.
Notice that in no decay is there interference
between the scalar $a$ couplings and the pesudoscalar $b$ couplings.

The same is true in the production mechanisms.
We find
\be
\sigma (g g \rightarrow h)
=
\frac{G_\mu \alpha_s^2}{512 \sqrt{2} \pi}
\left[
\left| \sum_q a\, A_{1/2} (\tau ) \right|^2
+
\left| \sum_q b\, A^A_{1/2} (\tau ) \right|^2
\right],
\ee
where the sums, which run over all quarks $q$,
are dominated by the triangle with top in the loop with,
depending on $\tan{\beta}$,
relevant contributions from the triangle with bottom in the loop.
Therefore, we use
\be
\frac{\sigma (g g \rightarrow h)}{\sigma^{\textrm{SM}} (g g \rightarrow h)}
=
\frac{|a_t\, A_{1/2} (\tau_t) + a_b\, A_{1/2} (\tau_b)|^2 +
|b_t\, A^A_{1/2} (\tau_t) + b_b\, A^A_{1/2} (\tau_b)|^2 }{
|A_{1/2} (\tau_t) + A_{1/2} (\tau_b)|^2}.
\ee
Similarly,
\be
\frac{\sigma_{VBF}}{\sigma^{\textrm{SM}}_{VBF}}
=
\frac{\sigma_{VH}}{\sigma^{\textrm{SM}}_{VH}}
= C^2,
\ee
and
\be
\frac{\sigma (b \bar{b} \rightarrow h)}{\sigma^{\textrm{SM}} (b \bar{b}  \rightarrow h)}
= a^2 + b^2.
\ee
Notice that these expressions hold for any model with the effective lagrangians of
Eqs.~\eqref{LY}-\eqref{LVV}.

Our results agree with those of Choi \textit{et.\/ al}
\cite{Choi:1999at,Choi:2001iu}.
For ease of reference,
we show their definitions in terms of ours:
\ba
F_{s f} (\tau ) &=& \tfrac{1}{2} A_{1/2}(\tau ),
\nonumber\\
F_{p f} (\tau ) &=& \tfrac{1}{2} A^A_{1/2}(\tau ),
\nonumber\\
F_1 (\tau ) &=& - A_{1}(\tau ),
\nonumber\\
F_0 (\tau ) &=& A_{0}(\tau ),
\ea
\be
C_i = \lambda,
\ee
\be
(c_\beta O_{2,i} + s_\beta O_{3,i}) = C,
\ee
and
\ba
g_{s f}^i &=& - (\sqrt{2} G_\mu)^{1/2} m_f a = - \frac{g}{2 M_W} m_f a,
\nonumber\\
g_{p f}^i &=& - (\sqrt{2} G_\mu)^{1/2} m_f b = - \frac{g}{2 M_W} m_f b.
\ea


\begin{thebibliography}{99}
\bibitem{Higgs}
Applying the same transformations to the real parts of the
neutral fields (and, thus, to all fields) would lead us into
the Higgs basis \cite{LS,BLS}:
%
\be
\left(
\begin{array}{c}
H_1\\
H_2
\end{array}
\right)
=
\left(
\begin{array}{cc}
c_{\beta} & s_{\beta}\\
- s_{\beta} & c_{\beta}
\end{array}
\right)
\left(
\begin{array}{c}
\phi_1\\
\phi_2
\end{array}
\right),
\ee
%
with
%
\be
H_1 =
\left(
\begin{array}{c}
G^+\\
\tfrac{1}{\sqrt{2}} (v + H^0 + i G^0)
\end{array}
\right),
\hspace{5ex}
H_2 =
\left(
\begin{array}{c}
H^+\\
\tfrac{1}{\sqrt{2}} (R_2 + i I_2)
\end{array}
\right).
\ee
%
The relation between the scalar fields is
%
\be
\left(
\begin{array}{c}
\eta_1\\
\eta_2\\
\eta_3
\end{array}
\right)
=
R_H
\left(
\begin{array}{c}
H^0\\
R_2\\
I_2
\end{array}
\right)
=
\left(
\begin{array}{ccc}
c_{\beta} & - s_{\beta} & 0\\
s_{\beta} & c_{\beta} & 0\\
0 & 0 & 1
\end{array}
\right)
\left(
\begin{array}{c}
H^0\\
R_2\\
I_2
\end{array}
\right),
\ee
%
leading to
%
\be
\left(
\begin{array}{c}
h_1\\
h_2\\
h_3
\end{array}
\right)
=
R
\left(
\begin{array}{c}
\eta_1\\
\eta_2\\
\eta_3
\end{array}
\right)
=
R\, R_H
\left(
\begin{array}{c}
H^0\\
R_2\\
I_2
\end{array}
\right).
\ee
%
The matrix $T = R_H^T R^T$ is needed in Eq.~(381) of Ref.~\cite{ourreview},
in order to compute the bounds from the oblique radiative corrections
in Eqs.~(388) and (393) of \cite{ourreview}.
\bibitem{J1}
Lavoura and Silva \cite{LS} have classified all
basis invariant measures of CP violation which can occur in the
scalar sector of the 2HDM.
In particular,
for $s_2=0$ and using \cite{Higgs},
we can show that the invariant in Eq.~(331) of Ref.~\cite{ourreview}
is proportional to
%
\be
J_1 \propto (m_1^2 -m_2^2)(m_1^2 -m_3^2)(m_2^2 -m_3^2)
\cos{(\beta -\alpha_1)} \sin^2{(\beta -\alpha_1)}
\cos{(\alpha_3)} \sin{(\alpha_3)}.
\ee
%
Therefore,
in general,
even if the lightest Higgs is CP even,
there can be CP violation in the mixing of the heavier two.
This source of CP violation only vanishes if the sine or cosine
of $\beta - \alpha_1$ or of $\alpha_3$ vanish,
or if the heavier two states are degenerate.
\end{thebibliography}
\end{document}